\documentclass[epj,referee]{svjour}
\usepackage{graphics}
\begin{document}
\sloppy
\title
{Polaron cross-overs and {\em d}-wave superconductivity in Hubbard-Holstein 
model}
\author{R. Ramakumar\thanks{{\em Present Address: Department of Physics and
Astrophysics, University of Delhi, Delhi-110007, India.}} \and A. N. Das}
\institute{Condensed Matter Physics Group, Saha Institute of Nuclear Physics,
1/AF Bidhannagar, Calcutta-700064, INDIA} 
\date{Received: 26 May 2004 / Revised version: 9 July 2004}
\abstract{We present a theoretical study of superconductivity 
of polarons in the
Hubbard-Holstein model.
A residual kinematic interaction proportional to the square of
the polaron hopping energy between polarons and phonons 
provides a pairing field for the polarons.
We find that superconducting instability in the $d$-wave channel
is possible with small transition temperatures which is maximum
in the large to small polaron cross-over region. 
An $s$-wave instability is found to be not possible when the
effective on-site interaction between polarons is repulsive.
} 
\maketitle
\section{Introduction}
\label{sec1}
Polaronic superconductivity has been a subject of interest
following discovery of superconductivity in  cuprate oxides
and some molecular conductors\cite{alex}.
A characteristic feature of these materials is that the
conduction electrons live in a narrow energy band.
It is also recognized that electron-phonon ({\em e-ph})
interaction is considerable in these materials.
The simplest model for narrow band electrons interacting with
local phonons is the Hubbard-Holstein model \cite{hub,hol}.
If the {\em e-ph} interaction is large enough, it can lead to the
formation of polarons.
Building on an earlier work \cite{beni},
Takada and Hotta recently studied \cite{takada} this
model in the strong
{\em e-ph} coupling limit where the polarons are small polarons.
For intermediate {\em e-ph} coupling, however, the polarons
may not be small polarons depending on {\em e-ph} coupling
strength and the ratio of the values of
the electronic hopping to the phonon energy.
In this paper we present a study of superconductivity in this range
to check the role of large-to-small polaron crossover on
superconductivity.
\section{Mean-field theory of superconductivity in the
Hubbard-Holstein model}
\label{sect2}
The Hubbard-Holstein (HH) model \cite{hub,hol} is:
\begin{eqnarray}
\displaystyle
\boldmath
H &=& -t\sum_{i{\delta}\sigma}c^{\dag}_{i\sigma}c_{i+\delta\sigma}
        +U\sum_{i}n_{i\uparrow}n_{i\downarrow}  \nonumber \\
&+&\sum_{i}b^{\dag}_{i}b_{i}+g\sum_{i\sigma}n_{i\sigma}(b^{\dag}_{i}+b_{i}).
\unboldmath
\end{eqnarray}
Here $t$ ($>\,0$) is the hopping energy between molecules at lattice site
$\boldmath {i}\,$ and its nearest-neighbor lattice sites
{\boldmath $i\,+\delta$},
$c_{i\sigma}$ ($c_{i\sigma}^{\dag}$)
is the annihilation (creation) operator for the electron with spin
$\sigma$ at  a lattice site $i$ and $n_{i \sigma}$ is the corresponding number
operator, $U$ is on-site Coulomb repulsion,
$b_i$ ($b_{i}^{\dag}$) is the phonon annihilation 
(creation) operator, 
and $g$ is the {\em e-ph} interaction strength.
All energies
are measured in terms of the phonon energy ($\hbar\omega$). 
In the calculations given below,
first we apply a Modified-Lang-Firsov
(MLF) transformation \cite{das} (which is the
original LF \cite{lf} transformation modified to include lattice deformations
on the sites nearest neighbor to the site where an electron resides)
to $H$ to convert it 
to the polaron representation,
then we eliminate the residual polaron-phonon ({\em pol-ph}) interactions 
to obtain an
effective interaction between polarons, and finally we do a 
mean-field theory of superconductivity specializing to the cases of
$s$ and $d$ wave order parameters in the case of a square lattice. 
Our main result is contained in Fig. 2, where it is shown that
the superconducting transition temperature in the $d$-wave channel
goes through its maximum in the large to small polaron cross-over
region (see Fig. 1). The details of our calculations are given
in the rest of this section and the conclusions are given in Sec. III.
\par
The application of
MLF transformation to $H$ 
leads to $H_{MLF}=e^{R}He^{-R}$, where $R$ is given by,
\begin{equation}
R= \lambda_{\circ}\sum_{i\sigma}n_{i\sigma}(b^{\dag}_{i}-b_{i})
+ \lambda_{1}\sum_{i\delta\sigma}n_{i\sigma}
      (b^{\dag}_{i+\delta}-b_{i+\delta})\,,
\end{equation}
$\lambda_0$ and $\lambda_1$ represent the lattice deformations
at the electron site and its nearest-neighbor sites, respectively.
They are treated as variational parameters
to be determined by minimization (with respect to $\lambda_{\circ}$
and $\lambda_{1}$) of ground state
energy of the transformed Hamiltonian. Using the above $R$, we obtain:
\begin{equation}
\displaystyle
e^{R}c_{i\sigma}e^{-R}=c_{i\sigma} 
       \,exp\left[-\lambda_{\circ}(b^{\dag}_{i}-b_{i})
         -\lambda_{1}\sum_{\delta}
        (b^{\dag}_{i+\delta}-b_{i+\delta})\right]
\end{equation}
and,
\begin{equation}
e^{R}b_{i}e^{-R}= b_{i}-\lambda_{\circ}\sum_{\sigma}n_{i\sigma}
  -\lambda_{1}\sum_{\delta\sigma}n_{i+\delta,\sigma}\,.
\end{equation}
The transformed Hamiltonian ($H_{MLF}$) is obtained as:
\begin{equation}
\displaystyle
H_{MLF} = H_{\circ} + H_{1}+H_{2}\,,
\end{equation}
where,
\begin{eqnarray}
\displaystyle
H_{\circ}&=&\sum_{i}b^{\dag}_{i}b_{i}
    -\epsilon_{p}\sum_{i\sigma}n_{i\sigma}
+U_{eff}\sum_{i}n_{i\uparrow}n_{i\downarrow}  
\nonumber \\ 
 &-&V_{1}\sum_{i\delta}\sum_{\sigma\sigma^{\prime}}
                n_{i\sigma}n_{i+\delta,\sigma^{\prime}}
     \nonumber \\
   &+& V_{2}\sum_{i\delta \delta^{\prime},
\delta\neq \delta^{\prime}}\sum_{\sigma\sigma^{\prime}}
n_{i+ \delta,\sigma}n_{i+\delta^{\prime},\sigma^{\prime}}\,,
\end{eqnarray}
\begin{equation}
H_{1}=-t\sum_{i\delta\sigma}c^{\dag}_{i\sigma}
            c_{i+\delta,\sigma}\,exp(X_{i}-X_{i+\delta})\,,
\end{equation}
and
\begin{eqnarray}
H_{2}&=&(g-\lambda_{\circ})\sum_{i\sigma}n_{i\sigma}(b^{\dag}_{i}+b_{i})
 \nonumber \\
&-& \lambda_{1}\sum_{i\delta\sigma}
    n_{i\sigma}(b^{\dag}_{i+\delta}+b_{i+\delta})\,.
\end{eqnarray}
Here,
\begin{equation}
\epsilon_{p}= \lambda_{\circ}(2g-\lambda_{\circ})
             -z\lambda^{2}_{1}\,,
\end{equation}
\begin{equation}
U_{eff}=U-2\epsilon_{p}\,,
\end{equation}
\begin{equation}
V_{1}=2(g-\lambda_{\circ})\lambda_{1}\,, 
V_{2}=\lambda^{2}_{1}\,,
\end{equation}
\begin{equation}
X_{i}=\lambda_{\circ}(b^{\dag}_{i}-b_{i})
      +\lambda_{1}\sum_{\delta}
        (b^{\dag}_{i+\delta}-b_{i+\delta})\,,
\end{equation}
and $z$ is the coordination number of a lattice site.
The Eq. (5) is the Hamiltonian of polarons. 
The polaron is much more massive than the original electron
since it is a composite particle consisting of an electron and
the associated lattice deformations. These 
polarons interact among themselves (through the interactions
$U_{eff}$, $V_1$ and $V_2$), and interact with the phonons
through off-diagonal phonon terms of $H_1$ and $H_2$. 
The phonons remain
unaffected by the MLF transformation. Notice also that the 
{\em pol-ph} couplings are much weaker than the original 
{\em e-ph} coupling ($g$). 
\par
Now we work using the variational phonon basis obtained
through the MLF transformation. The variational parameters
$\lambda_0$ and $\lambda_1$ are found out from a minimization
of the ground-state energy in the normal state with zero
phonon averaging. The values of $\lambda_0$ and $\lambda_1$
depend on $t/\omega$ and $g/\omega$ for a particular lattice
and not on $U$. With this variational basis we proceed with our
second order perturbation calculation by expanding 
the exponentials in
$H_{1}$ to obtain, up to $O(\lambda_{\circ}-\lambda_{1})$,
$\widetilde{H}$ valid for $(\lambda_{\circ}-\lambda_{1})\,<\,1$ as:
\begin{equation}
\displaystyle
\widetilde{H}= H_{\circ}-t_{p}\sum_{i\delta\sigma}
       c^{\dag}_{i\sigma}c_{i+\delta,\sigma}+H_{2}+H_{3},
\end{equation}
where,
\begin{eqnarray} 
H_{3}&=&-t_{p}\sum_{i\delta\sigma}
c^{\dag}_{i\sigma}c_{i+\delta,\sigma}\,
[(\lambda_{\circ}-\lambda_{1})
(b^{\dag}_{i}-b_{i}-b^{\dag}_{i+\delta}+b_{i+\delta})
\nonumber \\ 
&+&
\lambda_{1}\sum_{\delta^{\prime}\neq\delta}
(b^{\dag}_{i+\delta^{\prime}}-b_{i+\delta^{\prime}})
\nonumber \\
&-&
\lambda_{1}\sum_{\delta^{\prime}\neq-\delta}
(b^{\dag}_{i+\delta+\delta^{\prime}}
-b_{i+\delta+\delta^{\prime}})
]
\end{eqnarray}
and,
\begin{equation}
t_{p}=t\,exp\left[-(\lambda_{\circ}-\lambda_{1})^{2}
          -(z-1)\lambda^{2}_{1}\right]\,.
\end{equation}
In the above equation $H_{2}$ and $H_3$ describe the 
{\em pol-ph}
interactions.
On eliminating the {\em pol-ph} interaction terms and considering
second order processes only, we obtain (for $zt_{p}\,<\,1$),
\begin{equation}
\widetilde{\widetilde{H}}=H_{\circ}-t_{p}\sum_{i\delta\sigma}
       c^{\dag}_{i\sigma}c_{i+\delta,\sigma}+H_{4},
\end{equation}
where,
\begin{eqnarray}
\displaystyle
H_{4}&=&-2(g-\lambda_{\circ})^{2}\sum_{i}n_{i\uparrow}n_{i\downarrow}
       \nonumber \\
&+&2\lambda_{1}(g-\lambda_{\circ})
\sum_{i\delta}\sum_{\sigma\sigma^{\prime}}
                n_{i\sigma}n_{i+\delta,\sigma^{\prime}}
\nonumber \\
&-&\lambda^{2}_{1}
\sum_{i\delta \delta^{\prime}}\sum_{\sigma \sigma^{\prime}}
n_{i+\delta,\sigma}n_{i+\delta^{\prime},\sigma^{\prime}}
\nonumber \\
&-&t^{2}_{p}(\lambda_{\circ}-\lambda_{1})^{2}
\sum_{i\delta \delta^{\prime}}
\sum_{\sigma \sigma^{\prime}}
[c^{\dag}_{i-\delta^{\prime},\sigma^{\prime}}
c^{\dag}_{i\sigma}c_{i+\delta,\sigma}c_{i,\sigma^{\prime}}
\nonumber \\
&+&c^{\dag}_{i,\sigma^{\prime}}c^{\dag}_{i-\delta,\sigma}
c_{i,\sigma}c_{i+\delta^{\prime},\sigma^{\prime}}
-c^{\dag}_{i-\delta^{\prime},\sigma^{\prime}}
c^{\dag}_{i-\delta,\sigma}c_{i\sigma}c_{i\sigma^{\prime}}
\nonumber \\
&-&c^{\dag}_{i\sigma^{\prime}}c^{\dag}_{i\sigma}
c_{i+\delta,\sigma}c_{i+\delta^{\prime},\sigma^{\prime}}].
\end{eqnarray}
It may be noted that our study is restricted to the
region where $t_{p}(\lambda_{\circ}-\lambda_{1})$, 
$(g-\lambda_{\circ})$,
and $\lambda_{1}$ are much less than $1$ even though $g\,>\,1$.
This justifies our perturbation calculation. 
On simplification, confining to nearest neighbor singlet 
pairing terms, and neglecting terms $O(\lambda^{2}_{1})$ since
$\lambda_{1}$ is quite small (see Fig. 2), we get:
\begin{eqnarray}
\displaystyle
\widetilde{\widetilde{H}}&=&-t_{p}\sum_{i\delta\sigma}
       c^{\dag}_{i\sigma}c_{i+\delta,\sigma}
     -\epsilon_{p}\sum_{i\sigma}n_{i\sigma}
 +\widetilde{U}\sum_{i}n_{i\uparrow}n_{i\downarrow}
  \nonumber \\ 
&-&t^{2}_{p}(\lambda^{2}_{\circ}-2\lambda_{\circ} \lambda_{1})
\sum_{i\delta}\sum_{\sigma \sigma^{\prime}}
[
c^{\dag}_{i+\delta,\sigma^{\prime}}c^{\dag}_{i,\sigma}
c_{i+\delta,\sigma}c_{i\sigma^{\prime}}
\nonumber \\
&+&c^{\dag}_{i+\delta,\sigma^{\prime}}c^{\dag}_{i\sigma}c_{i+\delta,\sigma}
c_{i\sigma^{\prime}}
-c^{\dag}_{i+\delta,\sigma^{\prime}}c^{\dag}_{i+\delta,\sigma}
c_{i\sigma}c_{i\sigma^{\prime}}
\nonumber \\
&-&c^{\dag}_{i\sigma^{\prime}}c^{\dag}_{i\sigma}
c_{i+\delta,\sigma}c_{i+\delta,\sigma^{\prime}}
],
\end{eqnarray}
where $\widetilde{U}=(U-2g^{2})$.
The elimination of the {\em pol-ph} terms generates a pairing
field whose magnitude is $O(t^{2}_{p})$ which is rather small. 
Note that the field associated with the on-site pair-hopping 
processes (the last two terms
in $\widetilde{\widetilde{H}}$) is repulsive. 
The origin of superconductivity in this model, when $\widetilde{U}\,>\,0$, 
then is the field generated through a virtual emission and absorption of phonons
while the polarons hop from site to site.
The two-polaron process involved is one polaron hops to its 
nearest neighbor site
and excites a phonon at the target site and {\em another} polaron
on the target site absorbs the excited phonon and hops to the
site left by the first polaron.
These processes lead to the fourth and fifth terms in 
If the {\em e-ph} interaction is not sufficient
enough to lead to the formation of polarons, superconductivity occurs
in the usual way through the phonon exchange between electrons,
and the pairing field contains the full {\em e-ph} interaction.
Now we are at a stage where we can do a mean-field theory
of the superconductivity using $\widetilde{\widetilde{H}}$.
\par
On mean-field factorization of $\widetilde{\widetilde{H}}$, we obtain:
\begin{eqnarray}
\displaystyle
H_{MF}&=&-t_{p}\sum_{i\delta\sigma}
       c^{\dag}_{i\sigma}c_{i+\delta,\sigma}
     -(\epsilon_{p}-\widetilde{U}\frac{n}{2})\sum_{i\sigma}n_{i\sigma}
     -N\widetilde{U}\frac{n^{2}}{4}
 \nonumber \\
&+&\widetilde{\widetilde{U}}\sum_{i}
(A_{\circ}c_{i\downarrow}c_{i\uparrow}+h.c.)
-N\widetilde{\widetilde{U}}|A_{\circ}|^{2}
 \nonumber \\
&+&V_{tp}\sum_{i\delta}
(\Delta_{\delta}c_{i+\delta\downarrow}c_{i\uparrow}+h.c.)
-NzV_{tp}|\Delta_{\delta}|^{2}\,,
\end{eqnarray}
where
\begin{equation}
A_{\circ}=<c^{\dag}_{i\uparrow}c^{\dag}_{i\downarrow}>\,,
\end{equation}
\begin{equation}
\Delta_{\delta} 
=<c^{\dag}_{i+\delta\uparrow}c^{\dag}_{i\downarrow}>\,,
\end{equation}
\begin{equation}
\widetilde{\widetilde{U}}=\widetilde{U}-zV_{tp}\,,
\end{equation}
and
\begin{equation}
V_{tp}=-4t^{2}_{p}(\lambda^{2}_{\circ}-2\lambda_{\circ}\lambda_{1})\,.
\end{equation}
The above mean-field approximation is valid for $\tilde{U}$
less than the polaron band-width. For larger $\tilde{U}$ 
strong correlation effects like Hubbard sub-band formation
can occur. For recent studies of strong correlation effects
originating from large $U$ in Hubbard-Holstein model see Refs.
\cite{cast,hew}.
Converting to momentum space, we get:
\begin{eqnarray}  
H_{MF}&=&-t_{p}\sum_{{\bf k}\sigma}\xi({\bf k})
c^{\dag}_{{\bf k}\sigma}c_{{\bf k}\sigma}
-\left(\epsilon_{p}-\frac{\widetilde{U}n}{2}\right)
\sum_{{\bf k}\sigma}n_{{\bf k}\sigma}
 \nonumber \\
&+&\sum_{{\bf k}}
\left[\left(\widetilde{\widetilde{U}}A_{\circ}+V_{tp}A_{1}({\bf k})\right)
c_{-{\bf k}\downarrow}c_{{\bf k}\uparrow}+h.c.\right]
\nonumber \\
&-&N\widetilde{\widetilde{U}}|A_{\circ}|^{2}-NzV_{tp}|\Delta_{\delta}|^{2}\,,
\end{eqnarray}
where
\begin{equation}
\xi({\bf k})= 
\sum_{\delta}e^{i{\bf k.\delta}},\,
\end{equation}
\begin{equation}
A_{1}({\bf k})=\sum_{\delta}\Delta_{\delta}e^{i{\bf k.\delta}}\,,
\end{equation}
and $N$ is the number of lattice sites.
In the $s$-wave case, $\widetilde{U}$ term should be considered even if 
$\widetilde{U}\,>\,0$ since the extended $s$-wave  order parameter
has an on-site component, and this is suppressed by the on-site
repulsion $\widetilde{U}$. The $\widetilde{U}$ term does not affect $d$-wave
pairing. In studying superconductivity, we will be considering only
$\widetilde{U}\,>\,0$ case since
$\widetilde{U}\,<\,0$ case leads to an effective
attractive Hubbard model situation and this model is well
understood\cite{mic,met,beck}.
Now we will consider $s$ and $d$ wave pairing channels separately.
\subsection{s-wave channel}
\label{subsec}
In the $s$-wave case, the effective Hamiltonian is
\begin{equation}
\displaystyle
H_{s}=\sum_{{\bf k}\sigma}
    \tilde{\epsilon}({\bf k})c^{\dag}_{{\bf k}\sigma}c_{{\bf k}\sigma}
+\sum_{{\bf k}}(D_{s}({\bf k})
c_{-{\bf k}\downarrow}c_{{\bf k}\uparrow}+h.c.)+s_{\circ}\,,
\end{equation}
where, 
$\tilde{\epsilon}({\bf k})=-t_{p}\xi({\bf k})-\tilde{\mu}$,
$D_{s}({\bf k})=\widetilde{\widetilde{U}}A_{\circ}+V_{tp}A^{s}_{1}({\bf k})$,
$A^{s}_{1}({\bf k})=2\Delta_{\delta}\gamma_{s}({\bf k})$,
$\tilde{\mu}=\mu+\epsilon_{p}-\widetilde{U}n/2$,
$\gamma_{s}(k) = cos(k_{x}a)+cos(k_{y}a)$ for a square lattice
of lattice constant $a$, and $s_{\circ}=-(N\widetilde{U}n^{2}/4)
-n\widetilde{\widetilde{U}}|A_{\circ}|^{2}-NzV_{tp}|\Delta_{\delta}|^{2}$.
In the above we have introduced a chemical potential ($\mu$) to
fix the number density, and $A^{s}_{1}(k)$ is for the isotropic case.
The integral equation which determines
$D_{s}({\bf k})$ is obtained, using Green's functions, to be,
\begin{equation}
D_{s}({\bf k})=-\frac{1}{N}\sum_{{\bf k^{\prime}}}
V_{s}({\bf k\,k^{\prime}})
\frac{D_{s}({\bf k^{\prime}})}{2E_{s}({\bf k^{\prime}})}
tanh\left[\frac{\beta E_{s}({\bf k^{\prime}})}{2}\right]\,,
\end{equation}
where the pairing field $V_{s}({\bf k\,k^{\prime}})$ is
\begin{equation}
V_{s}({\bf k\,k^{\prime}})=
\widetilde{\widetilde{U}}+4z^{-1}V_{tp}
\gamma_{s}({\bf k})\gamma_{s}({\bf k^{\prime}}),\\
\end{equation}
$\beta=1/k_{B}T$, $k_{B}$ is the Boltzmann constant,
and the quasiparticle energy
$E_{s}({\bf k})=\sqrt{\tilde{\epsilon}^{2}({\bf k})+|D_{s}({\bf k})|^{2}}$.
Now, we have already mentioned that we will consider the non-trivial
case of $\widetilde{U}\,>\,0$. In that case $s$-wave pairing is not
possible since $V_{s}({\bf k\,k^{\prime}})$ is not attractive.
\subsection{d-wave channel}
\label{subsec}
In this case, we have
\begin{equation}
H_{d}=\sum_{{\bf k}\sigma}
    \tilde{\epsilon}({\bf k})c^{\dag}_{{\bf k}\sigma}c_{{\bf k}\sigma}
+\sum_{{\bf k}}(D_{d}({\bf k})
c_{-{\bf k}\downarrow}c_{{\bf k}\uparrow}+h.c.)+d_{\circ}\,,
\end{equation}
where
$D_{d}({\bf k})=V_{tp}A_{1}({\bf k})$,
and $d_{\circ}=-(NUn^{2}/4)-NzV_{tp}|\Delta_{\delta}|^{2}$.
Then the $d$-wave gap equation is obtained as
\begin{equation}
D_{d}({\bf k})=-\frac{1}{N}\sum_{{\bf k^{\prime}}}
V_{d}({\bf k\,k^{\prime}})
\frac{D_{d}({\bf k^{\prime}})}{2E_{d}({\bf k^{\prime}})}
tanh\left[\frac{\beta E_{d}({\bf k^{\prime}})}{2}\right]\,,
\end{equation}
where,
\begin{equation}
V_{d}({\bf k\,k^{\prime}})=4z^{-1}V_{tp}
\gamma_{d}({\bf k})\gamma_{d}({\bf k^{\prime}})\,,
\end{equation}
$E^{2}_{d}({\bf k})=\tilde{\epsilon}^{2}({\bf k})+|D_{d}({\bf k})|^{2}$,
and $\gamma_{d}(k)=cos(k_{x}a)-cos(k_{y}a)$ for a 2D-square-lattice 
case. The superconducting transition temperature $T^{d}_c$ is obtained from,
\begin{equation}
1=-\frac{1}{N}\sum_{{\bf k^{\prime}}}
V_{tp}
\frac{\gamma^{2}_{d}({\bf k^{\prime}})}
{2\tilde{\epsilon}({\bf k^{\prime}})}
\,tanh\left[\frac{\beta_{c}\tilde{\epsilon}({\bf k^{\prime}})}{2}\right]\,.
\end{equation}
where $\beta_{c}=1/k_{B}T^{d}_c$.
\par
We have numerically analyzed the above equations and the results are
shown in Figs. 1-2. 
The values of $t$ ($<\,0.5$) used are in the anti-adiabatic limit.  
In Fig. 1 we have
displayed the variation 
of the lattice deformation ($\lambda_{\circ}$ and $\lambda_{1}$)
as a function of $g$.
A smooth cross-over
from a large to a small polaron \cite{das} with increasing $g$
is seen in Fig. 1.  
\par
The variation of $T_c$ with $g$ (for $n=0.81$)
for superconducting instability in the $d$-wave 
channel is displayed in Fig. 2. The variation of
$T_c$ with $n$ has a bell shaped form centered at half-filling
similar to that found in Ref. \cite{takada}.  
Superconducting instability with small $T^{d}_{c}$ is
possible for intermediate values of {\em e-ph} coupling.
The small $T^{d}_{c}$ is due to small value of 
the pairing field [which is $O(t_{p}^{2})$].
The $T^{d}_{c}$ is seen to reach a maximum 
in the large to small polaron cross-over region.
\section{Conclusions}
\label{sec3}
In this work we presented  a study of superconductivity
in the anti-adiabatic limit of the Hubbard-Holstein model
considering the polaron formation. Our purpose was to study
superconductivity of the polarons 
in the large to small polaron cross over region.
We found that transition temperature for superconducting instability 
in the $d$-wave channel
goes through its maximum in the
large to small polaron cross-over regime. We also found that
$s$-wave pairing is not possible
for the case of $U\,>\,2g^{2}$. 
In the strong {\em e-ph} coupling limit
also, it was found \cite{takada} that $T^{d}_c$ is 
extremely small when $U\,>\,2g^{2}$.
In real materials there is inter-site
Coulomb repulsion and this will
drastically reduce \cite{inter} the $T^{d}_c$. So, we conclude
that,
when the mechanism considered here operates,
$T^{d}_c$ is very small both in the intermediate and strong coupling
limits of the {\em e-ph} interaction.
{}
\begin{figure}
\resizebox*{3.1in}{2.5in}{\rotatebox{270}{\includegraphics{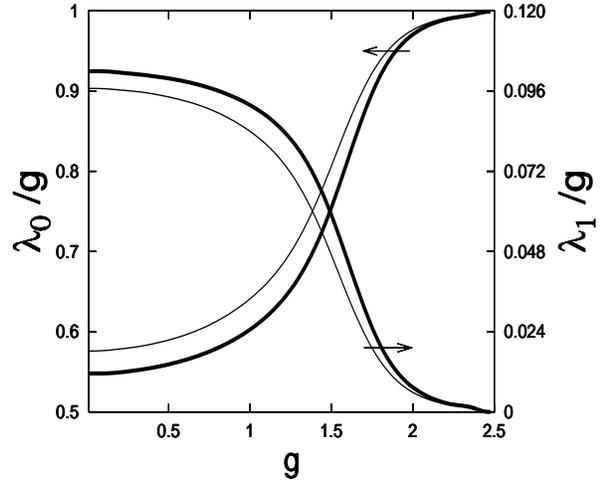}}}
\vspace*{0.5cm}
\caption[]{The variational parameters $\lambda_{0}$ and $\lambda_{1}$ for
$n\,=\,0.81$ and $t$: 0.35 (thick lines) and 0.30 (thin lines).}
\label{scaling}
\end{figure}
\begin{figure}
\resizebox*{3.1in}{2.5in}{\rotatebox{270}{\includegraphics{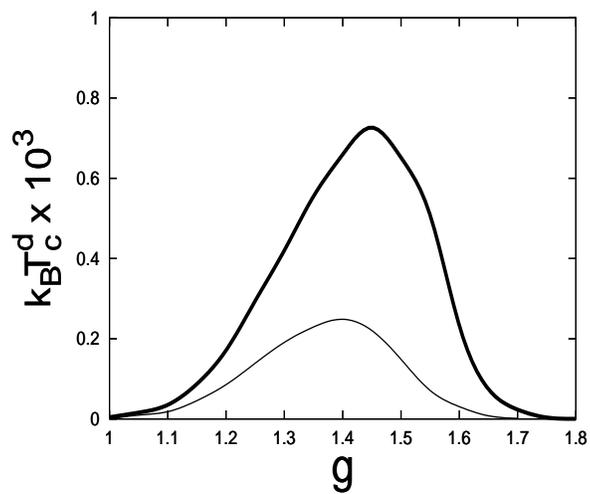}}}
\vspace*{0.5cm}
\caption[]{$T^{d}_{c}$ vs $g$ for $n\,=0.81$ and $t$: 0.35 (thick line)
and 0.30 (thin line).}
\label{scaling}
\end{figure}
\end{document}